# Architecture of monitoring elements for the network element modeling in a Grid infrastructure

A.Ciuffoletti, T.Ferrari
*INFN CNAF, Bologna, I-40127 ITALY*

A.Ghiselli, C.Vistoli
*INFN CNAF, Bologna, I-40127 ITALY*

Several tools exist that collect host-to-host connectivity measurements. To improve the usability of such measurements, they should be mapped into a framework consisting of complex subsystems, and the infrastructure that connects them. We introduce one such framework, and analyze the architectural implications on the network structure. In our framework, a complex subsystem consists of several computing facilities and the infrastructure that connects them: we call it a -monitoring domain-. The task of measuring the connectivity between -monitoring domains- is considered distinct from the activity of -storage- and -computing- elements. Therefore we introduce a new element in our topology: we call it -theodolite- element, since its function is similar to that of a transponder. Using these basic concepts, we analyze the architectural implications on the network structure: in a nutshell, if we want that -theodolites- serve as a reference, than the contribution to the relevant network metrics due to the -monitoring domain- infrastructure must be negligible with respect to contributions of the inter-domain infrastructure. In addition all -theodolites- of a -monitoring domain- must give an image of the inter-domain infrastructure that is consistent with that experienced by network applications. We conclude giving a running SQL example of how information about -monitoring domains- and -theodolites- could be organized, and we outline the application of such framework in the GLUE schema activity for the network element.

## 1. INTRODUCTION

Network service is a key element for distributed computing. A Grid computation requires three kinds of resources: Storage, Computing and communication resources then is appropriate to introduce a concept that reflects communication resources: in analogy with the Storage Entities (storage resources), and Computing Entities (computational resources), we opt for the term *Communication Entity*. Communication Entity has to be characterized in order to set up the best execution environment for the applications

## 2. COMMUNICATION ENTITY MODEL

The simplest definition of Communication Entity (in the following abbreviated as Communication) associates a Communication to each path in the grid. Although this option returns a fine grain image of the communication resources, we believe it is inappropriate for the following reasons:

- ?? it masks the fact that there are large LAN domains containing several grid resources having the same WAN connectivity;
- ?? it assumes that CEs and SEs are charged of network monitoring tasks;
- ?? It introduces the task of aggregating measures even for solving simple tasks: for instance, to compute a metric estimate between a CE and a SE in different domains, one would probably need to compose the metric estimate of the paths from the CE to the router in CE's domain, the one from the SE to the router in SE's domain, and the one between the two routers.

Each of the above facts contributes to render an image of the system that contains either unnecessary or poorly adherent with the needs of a grid monitoring architecture.

Therefore we opt for a model that does not bind Communication Entities to paths in the GRID, and gives an abstract view of the GRID and of its communication infrastructure. This drastically reduces the amount of information to store, avoids the overloading of the CE with monitoring capabilities, and may reduce the need of measurement aggregation.

### 2.1. A model of the Network resource in the GRID

We first introduce a new kind of entity, whose task is mainly that of measuring the performance. There are cases where this function is embedded in the Storage/Computing Entity functionality, others where this functionality is better associated to a distinct host. We call Theodolite Entity a host that measures network performance, using either active or passive tools.

We consider the Grid as partitioned into distinct Monitoring Domains (indicated in the following as Domains, capital D): a Domain is a set of Computing, Storage Entities, and Theodolite Entities. A Domain may reflect an existing organization, but may also indicate a part of a larger organization (a specialized laboratory), or an ``inter organizational" entity (a Grid backbone).

In order to give a precise definition of Domain, we need to introduce the concept of Connectivity. Connectivity is a metric that reflects the quality of communication through a link between two Entities, chosen among the Computing, Storage and Theodolite ones (also called Edge Entities). We understand that its definition is a work in progress, therefore we make just two wide-spectrum assumptions:

- ?? a higher Connectivity means better network performance

?? Connectivity can be used to define a (total) ordering among links.

In particular we do not assume that Connectivity has the mathematical features of a distance, or that the Connectivity of links having one edge in common can be aggregated in order to give the Connectivity of the overall path. Connectivity may have a dimension (e.g. kbytes/sec) or not (e.g. dB).

One basic property of a Domain is that the Connectivity between Edge Entities inside the Domain is (far) better than the Connectivity with Edge Entities in other Domains.

As for the deployment of the monitoring activity, some of the Edge Entities are Theodolite Entities: they run dedicated code that is used to compute network performance metrics, and in particular Connectivity.

The publication of such information, as well as that related to the resources available at CEs and SEs, is demanded to a specialized service, the GRIS.

## 2.2. Communication Definition.

Given the definition of Network Monitoring Domain, the definition of Communication Entity is consequential.

?? the Communication Entities of a Grid are all pairs of Network Monitoring Domains in the Grid

?? the connectivity associated to a Communication Entity corresponds to the Connectivity between two designated Theodolite Entities inside the two Network Monitoring Domains

We associate a Communication Element to each pair of Theodolites. A Theodolite may be designated for several Communications.

### 2.2.1. Some practical consequences of the above definitions

An immediate consequence is that the internal Connectivity of a Domain must be carefully monitored, and compared with external connectivity. The following properties should be enforced, either manually or automatically:

?? inside a Domain (the Connectivity of) all paths between Storage/Computing Entities and gateways is comparable to (the Connectivity of) all paths between Storage/Computing Entities and Theodolites. This guarantees that the cost for reaching an exit point is comparable for the Theodolites and for the Storage/Computing ones.

?? (the Connectivity of) paths between Theodolites and the outside is negligible with respect to (the Connectivity of) internal paths.

This guarantees that:

?? Theodolites estimate the connectivity between the internal Computing/Storage Entities and the Computing/Storage Entities in the target Domain;

?? Computing/Storage Entities that log the network performance of application runs (e.g. gridftp) may export their results as Communication measures.

From the above definition one may conclude that the generated graph is a complete mesh: each Domain measures its Connectivity with all other Domains. The cost of the monitoring activity therefore grows with the square of the number of Domains in the Grid. One would prefer that such cost would depend on the number of direct inter-Domain paths. Attaining such result depends on the existence of composition rules for ``Connectivity".

## 2.3. The relational tables

The relations needed to address the above structure are the following:

?? a domain relation between Computing/Storage Entities and Domains

| Field | Type | Null | Key | default | extra |
|---|---|---|---|---|---|
| Entity | varchar(16) | | | | |
| Domain | varchar(16) | | | | |

?? a correspondence between couples of Domains and couples of Theodolites

| Field | Type | Null | Key | default | extra |
|---|---|---|---|---|---|
| DomainA | varchar(16) | | | | |
| Domain B | varchar(16) | | | | |
| TheodoliteA | varchar(16) | | | | |
| TheodoliteB | varchar(16) | | | | |

Computing, Storage and Thedolite Entities can be identified by their IP address. Network Monitoring Domains might be identified by an arbitrary identifier, probably not an IP address or mnemonic. Communication Entities are represented by couples of Domains, and do not appear explicitly in the relations.

This is a MySQL query for a Consumer that asks for the NetworkPacketLoss between a Computing Entity C2, and a Storage Entity S3:

```
select @Dc:=Domain from domain where Entity='C2';
select @Ds:=Domain from domain where Entity='S3';
select measure
from
NetworkPacketLoss,theodolite
where
theodolite.DomainA=@Dc &&
theodolite.DomainB=@Ds &&
NetworkPacketLoss.TheodoliteA =
theodolite.TheodoliteA &&
NetworkPacketLoss.TheodoliteB =
theodolite.TheodoliteB;
```

This is a MySQL query for a Consumer that asks for the NetworkPacketLoss between a Computing Entity C2, and any Storage Entity:

```
select @Dc:=Domain from domain where Entity='C2';
```

```
select domain.Entity,measure
from
NetworkPacketLoss,theodolite,domain
where
domain.Entity LIKE 'S%' &&
theodolite.DomainA=@Dc &&
theodolite.DomainB=domain.Domain &&
NetworkPacketLoss.TheodoliteA =
theodolite.TheodoliteA &&
NetworkPacketLoss.TheodoliteB =
theodolite.TheodoliteB;
```

This is a MySQL query for a Consumer that asks for the Storage Entity with lowest PacketLoss with respect to Computing Entity C2:

```
select @Dc:=Domain
from
domain
where Entity='C2';
```

Create temporary table PacketLossToStorage:

```
select domain.Entity,measure
from
NetworkPacketLoss,theodolite,domain
where
domain.Entity LIKE 'S%' &&
theodolite.DomainA=@Dc &&
theodolite.DomainB=domain.Domain &&
NetworkPacketLoss.TheodoliteA =
theodolite.TheodoliteA &&
NetworkPacketLoss.TheodoliteB =
theodolite.TheodoliteB;
select Entity from PacketLossToStorage order by
measure limit 1;
drop table PacketLossToStorage;
```

This is the access sequence for a (Producer) Theodolite T4 running an active monitoring tool, that wants to discover its partners:

```
select theodolite.TheodoliteB from theodolite
where theodolite.TheodoliteA='T4';
```

## 2.4. Communication entity and the GLUE information system

The communication entity model has been proposed and discussed in the GLUE collaboration[2] The Grid Laboratory Uniform Environment (GLUE) is a collaboration effort between Grid computing related projects with the aim to produce common core grid services to enable interoperability between various grid infrastructures. Since the network is a world wide basic resource, and it is important that the domain based network model is agreed in a very large communities to be useful in a grid distributed environment. The Glue agreement is then very important to be fulfilled to give grid-aware application network monitoring info available in a multiple grid domains infrastructure.

The conceptual model of the communication entity under discussion is the following:

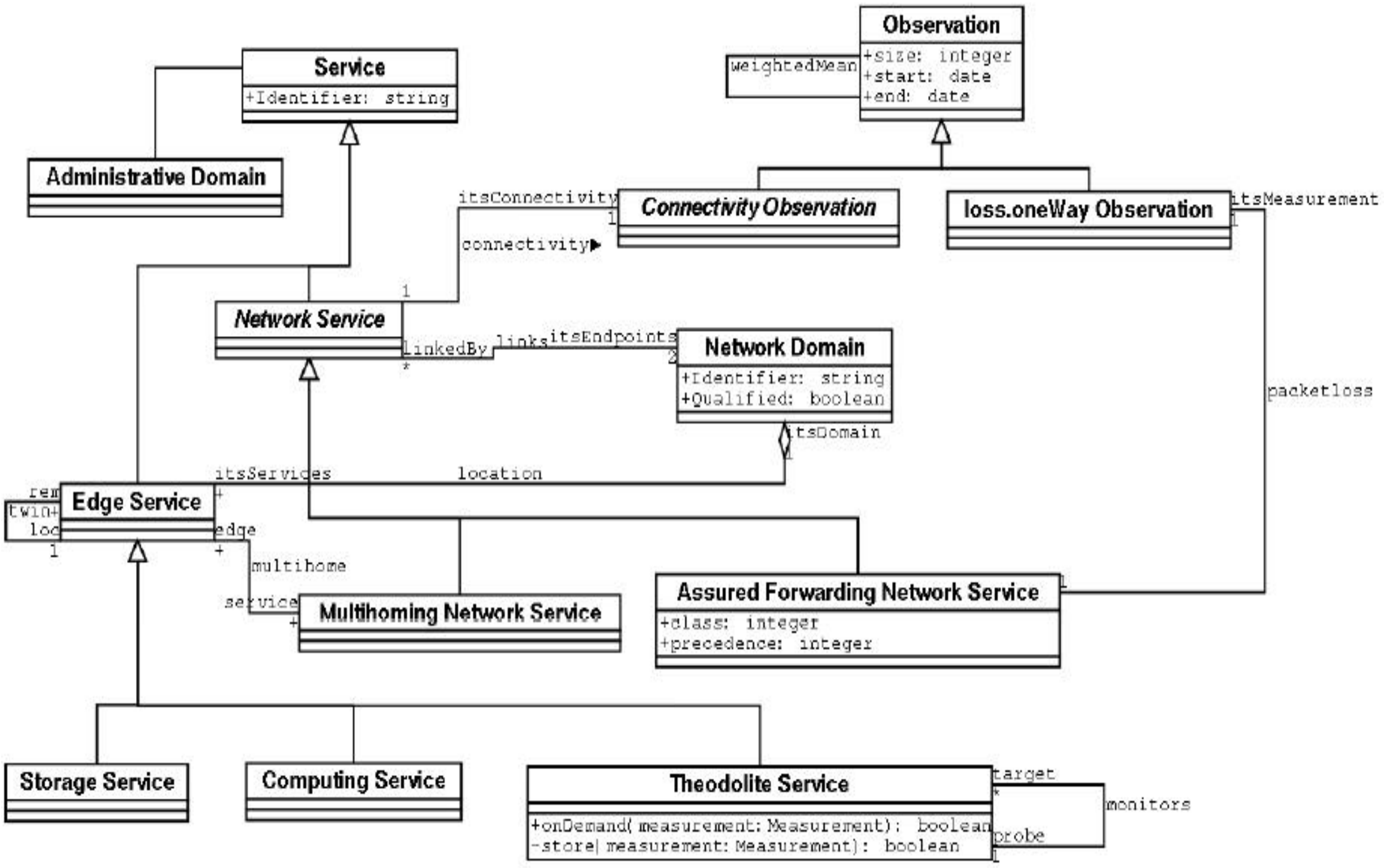

Figure 1: UML schema definition

## 2.5. Conclusions

Grid services and grid aware applications need to know network connectivity characteristics in order to make proper choices between the grid resources. A complete NxN measurement collection between the computing and storage resources is not scalable and a new model has to be defined. This document describes a model to partition the grid into network domains on the basis of a new parameter called 'connectivity': it assumed that the 'connectivity' is measured within the domain is always better than the 'connectivity' measured between domains. To collect the monitoring metrics of the communication entities between domains, a theodolite service has been defined. All the measured metrics are collected in grid information system based on a relational database technology.

## Acknowledgments

The authors wish to thank Sergio Andreozzi and all the Glue Collaboration.

Work supported by European DataTAG Project[1]